\documentclass[aps,pre,reprint,floatfix,groupedaddress]{revtex4-2}
\usepackage{cmap}
\usepackage[T2A]{fontenc}
\usepackage[utf8x]{inputenc}
\usepackage[russian,english]{babel}
\usepackage{graphicx}
\usepackage{hyperref}
\usepackage{amsmath,amssymb}
\usepackage{algorithmicx}   
\usepackage{algpseudocode}  
\DeclareMathOperator{\indeg}{deg^-}
\DeclareMathOperator{\outdeg}{deg^+}

\begin{document}

\title{Hierarchical crack patterns: Identification of crack generations}

\author{Yuri Yu. Tarasevich}
\email[Corresponding author: ]{tarasevich@asu-edu.ru}
\affiliation{Laboratory of Mathematical Modeling, Astrakhan Tatishchev State University, Astrakhan, 414056, Russia}

\author{Andrei S. Burmistrov}
\email{ksairen10@mail.ru}
\affiliation{Laboratory of Mathematical Modeling, Astrakhan Tatishchev State University, Astrakhan, 414056, Russia}

\author{Andrei V. Eserkepov}
\email{dantealigjery49@gmail.com}
\affiliation{Laboratory of Mathematical Modeling, Astrakhan Tatishchev State University, Astrakhan, 414056, Russia}

\date{\today}

\begin{abstract}
Identifying crack generations from microscopic images of hierarchical crack patterns is challenging due to the lack of temporal information and sensitivity to image boundaries. Existing algorithms often fragment individual cracks or lose stability when the observed fragment is shifted. We propose a method that reduces the classification problem to topological sorting of a directed acyclic graph (descendant$\to$parent), built from T-junctions and nearly collinear edges. Sequential removal of leaf vertices assigns generation numbers starting from the youngest. On 100 computer-generated networks, our method correctly classifies $\approx 70$\% of cracks at a window size of only three mean edge lengths, whereas a conventional approach that starts from primary cracks drops nearly to zero. The classification is highly stable against reasonable shifts of image boundaries but remains limited to strictly hierarchical networks.
\end{abstract}

\maketitle

\section{Introduction\label{sec:intro}}
Crack patterns are ubiquitous in the world around us~\cite{Goehring2015,Gupta2025}. These crack patterns often exhibit a hierarchical structure~\cite{Bohn2005,Schweiss2026}. Interest in crack patterns is driven, in particular, by the possibility of their use in modern technologies. For instance, crack-template-based (CTB) transparent conducting films (TCFs) are widely used in various devices: transparent heaters, shielding devices, electroluminescent devices, solar cells, smart windows, etc.~\cite{Cama2025a,Cama2025}.

When it comes to hierarchical crack patterns (HCPs), a distinction should be made between temporal and spatial hierarchies~\cite{Bohn2005}. Temporal hierarchy reflects the sequence of occurrence of the cracks that form the network. Naturally, to determine the temporal hierarchy, knowledge of the dynamics of the pattern formation is necessary. However, frequently only an image of the final pattern is available, thus, only the spatial hierarchy of cracks can be determined. Generally speaking, the temporal hierarchy of crack occurrence cannot be reconstructed from the image of the final HCP.

In the case of macroscopic HCPs, where the image represents the entire pattern and the width of the various cracks is easily measured, identifying relative crack generations poses no particular problem. The situation changes drastically for microscopic HCPs, such as those used as templates in the production of TCFs. In this case, the image represents only a tiny fraction of the entire HCP, while the image resolution is often insufficient to reliably measure crack widths, making this quantity difficult to use in the identification of crack generations. But the greatest difficulty is that it is impossible to reliably determine from an image of a network fragment which crack in the entire network arose first. This makes it impossible to construct a natural hierarchy from ancestor to descendants. The task is more akin to that of a person trying to reconstruct a family tree while knowing only a couple of recent generations of ancestors, and lacking the documents needed to identify the tree’s root. A classification of the spatial hierarchy of cracks is expected to be insensitive to reasonable shifts in the boundaries of the available image.

In the case of crack patterns with observable hierarchical structure, newly formed cracks connect to older pre-existing cracks at an angle close to $90^\circ$~\cite{Bohn2005,Lazarus2011,Goehring2013,Pauchard2020,Guan2025}, that is, crack junctions belong to $T$-type. If, after digitizing the image and transforming it into a graph embedded in the plane, the angle between a pair of adjacent edges is close to $180^\circ$, then such edges are considered to be successive segments of the same crack.

To identify crack generations in HCPs, a simple algorithm was proposed in which identification of crack generations starts from one or more cracks taken as roots~\cite{Bohn2005}. According to~\cite{Bohn2005}, primary cracks do not themselves connect directly to any other cracks; their ends are outside the observation window. Cracks that terminate on primary cracks are called secondary cracks. In general, a crack of order $n$ terminates at least at one of its ends at a crack of $(n-1)$-th order.

\begin{figure}[!htb]
 \centering
 \includegraphics[width=\columnwidth]{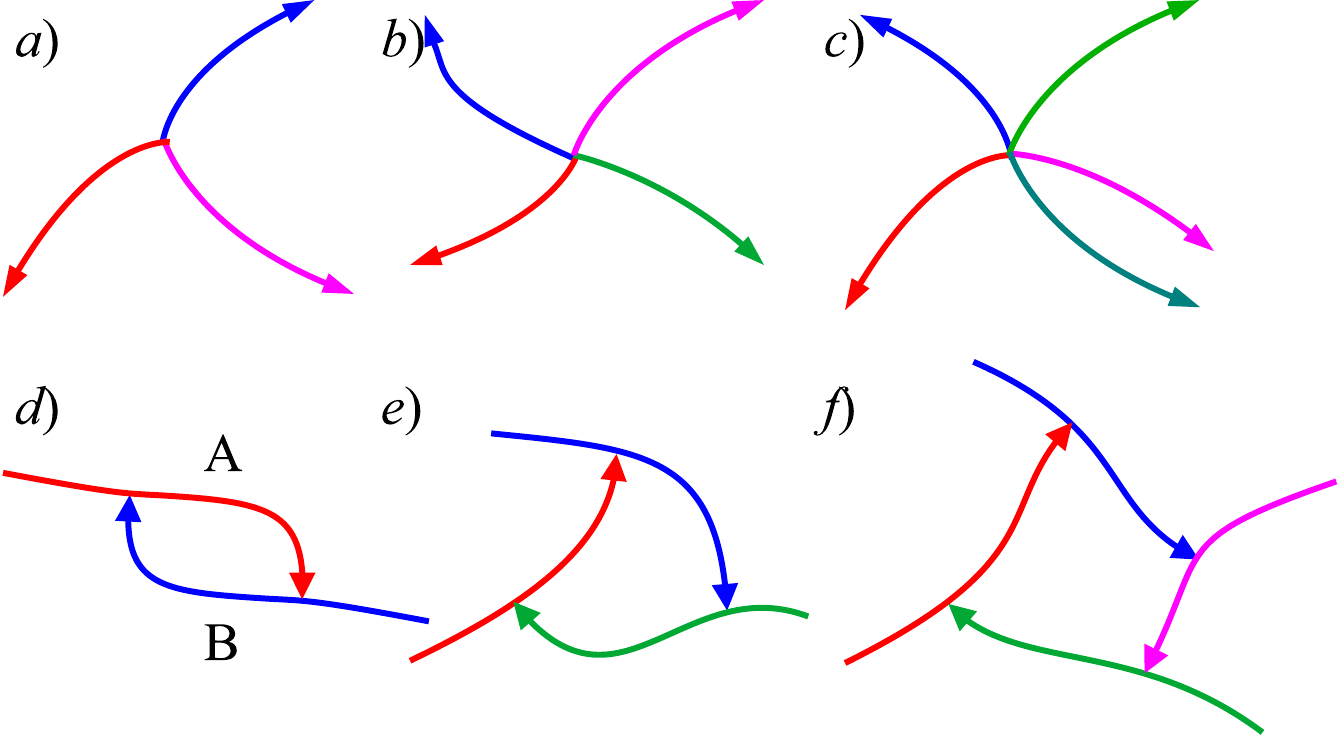}
 \caption{Examples of situations where the classification~\cite{Bohn2005} has to fail: a)--c)~stars, d)--f)~loops.}\label{fig:stars-and-loops}
\end{figure}

Classification~\cite{Bohn2005} was claimed by the authors to fail in two cases, viz., when (i) a 3-pointed star of cracks is formed at a defect of the layer and when (ii) three cracks form a loop.
Strictly speaking, the classification does not work for stars with any number of arms and for loops formed by any number of cracks (Fig.~\ref{fig:stars-and-loops}). Note that Fig.~\ref{fig:stars-and-loops}d exhibits an example of the so-called \emph{en passant} cracks~\cite{Kranz1979,Goehring2015}. All cracks forming a star or a loop obviously belong to the same generation. Indeed, the presence of loops and/or stars indicates that the network is not fully hierarchical, since a `descendant $\to$ ancestor' relationship cannot be established in these two cases. For example, in the case of a loop formed by two cracks propagating towards each other (en passant) (Fig.~\ref{fig:stars-and-loops}d), crack A enters crack B, so crack B is the ancestor of crack A, but crack B itself enters crack A, so crack A is the ancestor of crack B, leading to an unresolvable logical contradiction. Similar, but more involved, reasoning applies to loops formed by a larger number of cracks. In the case of a three-pointed star (Fig.~\ref{fig:stars-and-loops}a), for example, we have a $Y$-shaped connection of cracks, which is typical of non-hierarchical networks and indicates that all three cracks began to develop simultaneously from a common nucleation point.

%

To eliminate this drawback, an algorithm based on an auxiliary directed edge graph was proposed in~\cite{Perna2011}, along with its program code. Classification starts from the crack designated as the root. The method successfully solves the loop and star problems and, according to the authors, correctly classifies edges in cases where the approach of~\cite{Bohn2005} must fail. However, it is considerably more complex, sometimes erroneously assigns a lower order to some edges, and does not provide reliable classification of crack generations in certain situations.

In~\cite{Kumar2021}, alternative classifications were proposed. One of them is based on crack width analysis, while another is based on the analysis of angles between cracks. Unfortunately, this approach leads to fragmentation of cracks: parts of the same crack can be assigned to different generations, which is physically meaningless, because a crack is a single structural element that cannot belong to two different temporal or hierarchical generations simultaneously. According to~\cite{Kumar2021}, there is a regularity between the width and total length of cracks of different generations, and junctions of only $T$-type and $Y$-type were observed~\cite{Kumar2021}. The crack width distributions presented in~\cite{Kumar2021} allow us to roughly estimate the ratio of the width of first-, second-, and third-order cracks as $1:0.8:0.6$.
We highlight that different groups use different approaches to classify primary and secondary cracks, even when the classification is based on crack width; for example, the method presented in \cite[Fig.~4]{Voronin2022} differs from those in both~\cite{Bohn2005} and~\cite{Kumar2021}.

Note that the classifications~\cite{Bohn2005,Perna2011}, as well as the angle-based classification~\cite{Kumar2021}, interpret brickwork as a hierarchical network, although it is not one.
On the one hand, presence of $T$-junctions is not a sufficient condition for a network to be hierarchical. On the other hand, self-affinity of artificial networks~\cite{Fanfoni2025} does not ensure the presence of $T$-junctions which are typical for the real-world hierarchical crack patterns.

A classification of cracks was proposed based on the time dynamics~\cite{Nahlawi2006}, i.e. on the observed sequence of their occurrence. Primary cracks are the set of cracks that occur first, excluding cracks that occur between any two cracks that have formed. Secondary cracks are the second set of cracks that occur between two primary cracks. Tertiary cracks (if they occur) are the third set of cracks that occur between a primary and secondary crack or between two secondary cracks. This classification has been applied in a number of papers, e.g.~\cite{Tang2011,Tian2023,Yang2025b}. Indeed, the appearance of a crack is considered as an instantaneous event, rather than as a process developing in time.

Presence of crack generations requires that time is divided into some intervals, the boundaries of which are, generally speaking, quite arbitrary. The occurrence and development of a crack should occur within one interval, which is not the case: a crack can grow and expand during the entire drying time of the sample. Moreover, when crack generations are discussed, continuous time is replaced with discrete time: primary cracks are completely formed at the first temporal step, secondary cracks are formed at the second temporal step, etc. In most cases, only the final pattern of cracks is available for a study; the temporal dynamics of occurrence, propagation, development, and interaction of cracks is unknown.

Classification of cracks into different orders of a hierarchy can be done more or less unambiguously when one can observe cracking in real time or when one has available the entire image as opposed to a small subsection of the complete pattern. A major open question surrounds the accurate and unambiguous classification of cracks according to their hierarchy from photographs of such patterns, since these representations typically provide information only about a subpart of the sample at a fixed moment in time. It would be extremely helpful to find a rigorous method for crack classification when information about only a small part of the whole pattern is available for a single moment in time. Given that the widths of cracks are difficult to measure accurately and that differences in widths are frequently barely noticeable, the crack width is unlikely to be a promising quantity for such a classification.

\begin{figure}[!htb]
 \centering
\includegraphics[height=\columnwidth,angle=90]{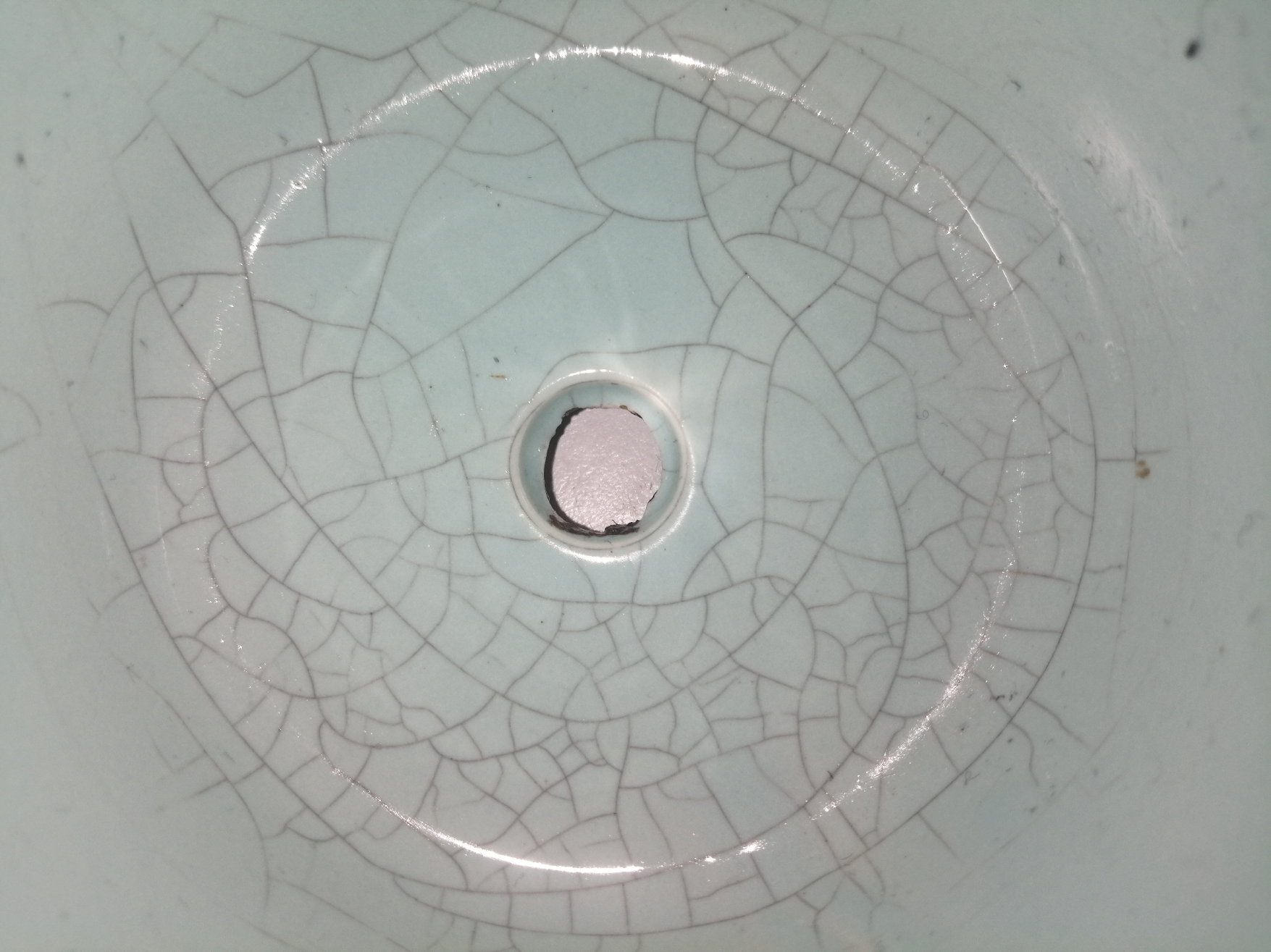}\hfill
 \caption{Cracked ceramic glaze on the internal surface of the plant pot. The hierarchical nature of the crack pattern is clearly visible.}\label{fig:pot}
\end{figure}

Crack patterns can be observed not only in simply connected domains, but also in multiply connected domains (see, e.g., Fig.~\ref{fig:pot}). When a domain is multiply connected, any deformations of the internal boundary cannot change a given crack hierarchy. A nucleation point can be treated as a point boundary. From the topological point of view, such a point defect can be stretched to a hole, which, indeed, forms an internal boundary. On the other hand, an internal boundary can be reduced to a single point. In particular, a loop (Fig.~\ref{fig:stars-and-loops}d--f) can be shrunk to a single point. Hence, stars and loops can be considered as two different instances of a common problem.

The aim of this work is to develop a method for classifying cracks in an image of a fragment of an HCP, which would have some stability with respect to displacements of the boundaries of the available image. The algorithm modification we propose is designed to address a fundamental flaw of the classification by angles~\cite{Kumar2021}: it fragments individual cracks, assigning different parts of the same crack to different generations. Our modification switches from junction analysis to crack analysis, preserving the key idea: in a T-shaped crack junction, the leg generation number is greater than the crossbar generation number.

The remainder of this paper is organized as follows. Section~\ref{sec:methods} describes the algorithm and simulation methods used. Section~\ref{sec:results} presents computer simulation results. Section~\ref{sec:concl} summarizes the main findings.

\section{Methods\label{sec:methods}}
\subsection{Preliminaries}
The image of the HCP is assumed to be already digitized and converted into a planar geometric graph embedded in the plane. We will assume that (i) this graph is not a tree, since the classification of generations on the tree is straightforward; (ii) there is only one component; (iii) the degree of any vertex does not exceed 3, meaning there are no multi-arm stars; (iv) there are no loops formed by two or more cracks. The last remark assumes that the network is completely hierarchical: for any contact between cracks, a child--parent relationship can be established.

The key concept for the following discussion is a crack, that is, a chain of successive edges that forms a continuous curve without sharp bends. Adjacent edges are considered to belong to the same crack only if the angle between them is close to $180^\circ$. We assume that the acceptable deviation should not exceed $30^\circ$, i.e., the minimum angle between edges lies in the range $[150^\circ,180^\circ]$, which, of course, is a matter of agreement. Nevertheless, this choice can be justified: Y-shaped connections, typical of non-hierarchical crack patterns, are characterized by angles close to $120^\circ$; the angle $150^\circ$ lies exactly in the middle between the typical value for HCPs and the typical value for non-hierarchical ones. In the extreme case, a crack can consist of a single edge if this edge has T-shaped connections with other cracks on both sides, or if one end of the edge is a T-shaped connection, and the other end terminates at the image boundary.

The algorithm's concept is based on the assumption that cracks, that is, chains of consecutive edges, can be defined uniquely; the ``child -- parent'' relationship can always be established unambiguously.
In general, this is not the case: in the presence of X-type connections, the identification of chains is ambiguous. In such cases, for example, one can assume that four cracks connect at a node, or, alternatively, one can assume that there is a single continuous crack with other cracks adjoining it on both sides.
Additional problems arise in the presence of stars, such as Y-shaped connections, and loops. For this reason, the restriction was introduced above that the degree of any vertex does not exceed 3 and that there are no loops.

In our study, the main data structure is a directed acyclic graph (DAG) in which the vertices correspond to cracks, while the arcs are directed from the child to the parent. Note that arcs are usually directed from parents to descendants; however, in our case, this choice is problematic because parents are far from always reliably known, whereas leaves (cracks with no descendants) are often known with certainty. For each T-shaped crack junction where a leg abuts a crossbar, an arc is added to the DAG from the vertex corresponding to the leg to the vertex corresponding to the crossbar. Generally speaking, an HCP is not a tree, however, it is convenient to call `leaves' those cracks that have no descendants ($\indeg(v)=0$), and to call `roots' those cracks which have no parents ($\outdeg(v)=0$). The main method is topological sorting, that is, ordering the vertices of a DAG according to a partial order defined by the edges of the DAG on the set of its vertices.

\subsection{Algorithm}

\begin{enumerate}
\item Find all cracks, that is, chains of consecutive edges with angles between them $\alpha \in [150^\circ; 180^\circ]$.
\item Find all leaves, that is, cracks that have no descendants:
\begin{enumerate}
\item both ends of a crack abut other cracks, forming T-junctions;
\item one end of a crack abuts another crack, forming a T-junction, and the node corresponding to the second end of the crack has degree 1.
\end{enumerate}
These leaves are assumed to belong to the youngest generation, hence, they have the highest order in the hierarchy. If there are no leaves, then the graph is a special type of graph, which requires a different processing algorithm; stop processing.
\item Starting from the leaves, a DAG is constructed whose vertices correspond to cracks and whose arcs are directed from the descendant to the ancestors.
\item In the resulting DAG, vertices with in-degree 0 ($\indeg(v)=0$) are leaves. We denote the level of the leaves by $n=0$. Here $n$ denotes the shortest distance in generations from the current crack to leaves.
\item All leaves along with their corresponding arcs are removed from the DAG and $n$ is incremented by one, $n \leftarrow n+1$. After removing all leaves, the in-degrees of some vertices in the remaining part of the DAG become equal to 0, $\indeg(v)=0$, that is, the vertices will become leaves.
\item Repeat the previous step until all vertices are removed from the DAG.
\end{enumerate}
Below this verbal description is presented in pseudocode form.
\begin{algorithmic}[1]
\State Find all cracks $C$ as chains of consecutive edges with angles $\alpha \in [150^\circ; 180^\circ]$ between them.
\State Determine adjacency relations between cracks.
\State $L \gets \varnothing$ \Comment{leaves (cracks without descendants)}
\ForAll {$c \in C$}
    \If {(both ends of $c$ are T-junctions) \textbf{or}
         (one end is a T-junction \textbf{and} the other end's node has degree 1)}
        \State $L \gets L \cup \{c\}$
    \EndIf
\EndFor
\If {$L = \varnothing$}
    \State \textbf{stop} \Comment{special graph type; requires different processing}
\EndIf
\State Build a directed acyclic graph $G = (V, A)$, where $V = C$ and arcs are directed from descendant to ancestor.
\State $n \gets 0$ \Comment{current level}
\State $R \gets V$ \Comment{vertices not yet removed}
\While {$R \neq \varnothing$}
    \State $L_n \gets \{ v \in R \mid \indeg_G(v) = 0 \}$ \Comment{vertices with in-degree 0}
    \ForAll {$v \in L_n$}
        \State $\mathrm{level}[v] \gets n$
        \State Remove $v$ and all outgoing arcs from $G$
    \EndFor
    \State $R \gets R \setminus L_n$
    \State $n \gets n + 1$
\EndWhile
\State \Return $\mathrm{level}$ \Comment{hierarchy level for each crack}
\end{algorithmic}

In such a way, the problem of identifying crack generations has been reduced to the well-known problem of topological sorting of a DAG, which in our case is carried out using an algorithm for sequentially removing leaves, i.e., vertices with in-degree 0, $\indeg(v)=0$, which is essentially a version of the well-known algorithm~\cite{Kahn1962}.
By contrast, the algorithm~\cite{Bohn2005} can be represented as sequentially finding all vertices with out-degree 0 ($\outdeg(v)=0$) and then removing them from the DAG.

Let there be a connected hierarchical crack network.
Let there be a fragment of this network to which a directed acyclic graph (DAG) is associated: the vertices are cracks, and the edges are directed from the younger generation to the older (from child to parent).

Then the minimum distance in generations from a given crack to a leaf (a crack with no children) equals the length of a shortest path from some leaf to the given crack, computed on this DAG using a standard shortest-path algorithm for directed acyclic graphs (for example, dynamic programming in topological order). This distance coincides with the true minimum distance in the full network provided that the fragment contains all cracks from which the given crack is reachable (i.e., all its descendants, including all leaves of its subtree).

\subsection{Analysis of the stability of crack classification with respect to image size}

To test the stability of the two crack classification methods under consideration with respect to image size, 100 hierarchical crack networks were used, which were generated on a computer using the simulation algorithm~\cite{Tarasevich2026}. For each network, cracks were classified using both methods, namely the method described in this paper and the method~\cite{Bohn2005}. For each network, a square fragment of size $w \times w$ was then extracted from the entire network. The cracks in this fragment were classified, and the fraction of cracks whose order matched that obtained for the entire network, $q$, was determined. The data for each window size were averaged over all networks. Since the hierarchical crack network has a natural characteristic length dependent on crack concentration, namely, the average edge length in the corresponding planar graph, $\langle l \rangle$, it is natural to use this quantity as a unit of length.

\section{Results\label{sec:results}}

\begin{figure*}
 \centering
 \includegraphics[width=\columnwidth]{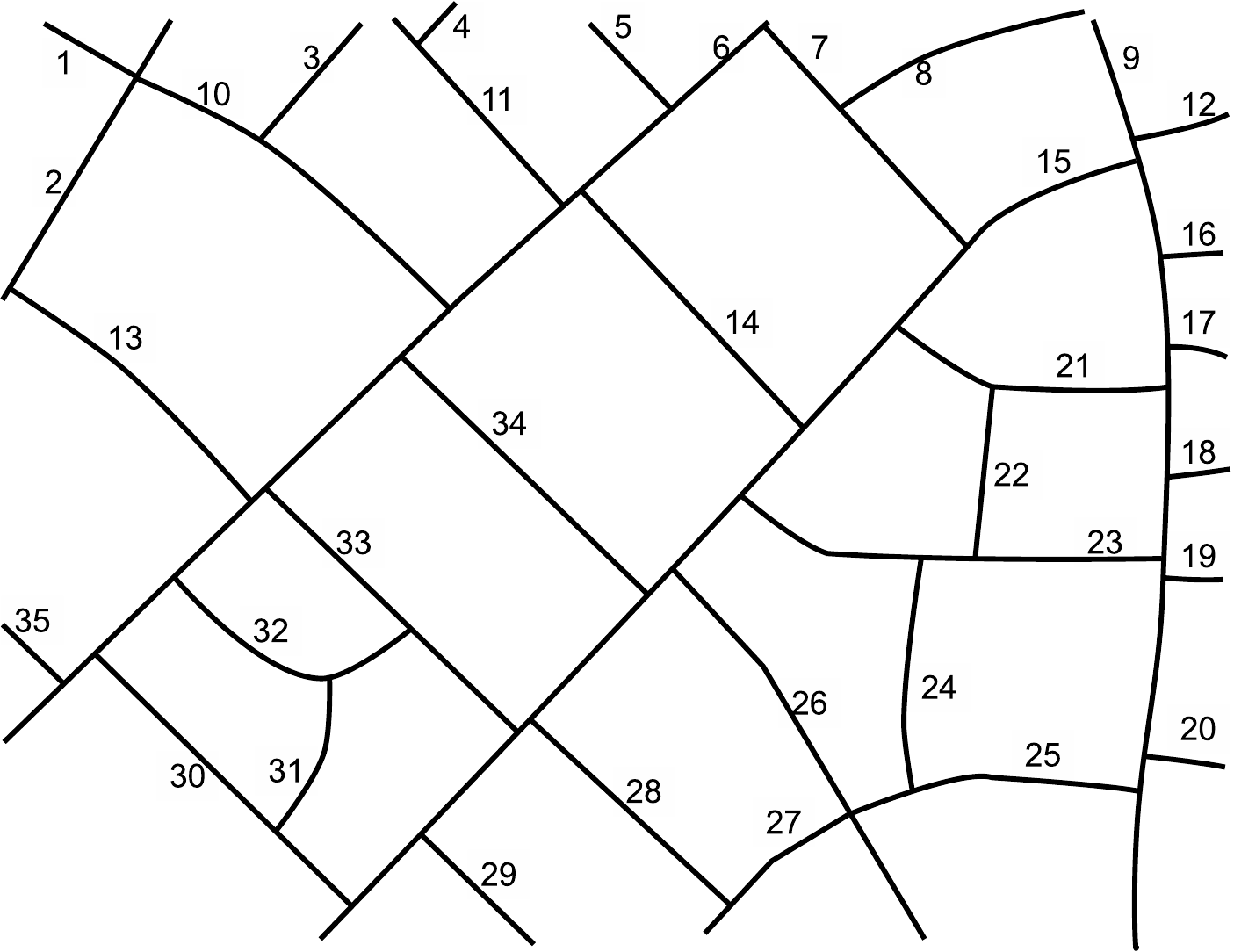}\qquad\includegraphics[height=0.77\columnwidth]{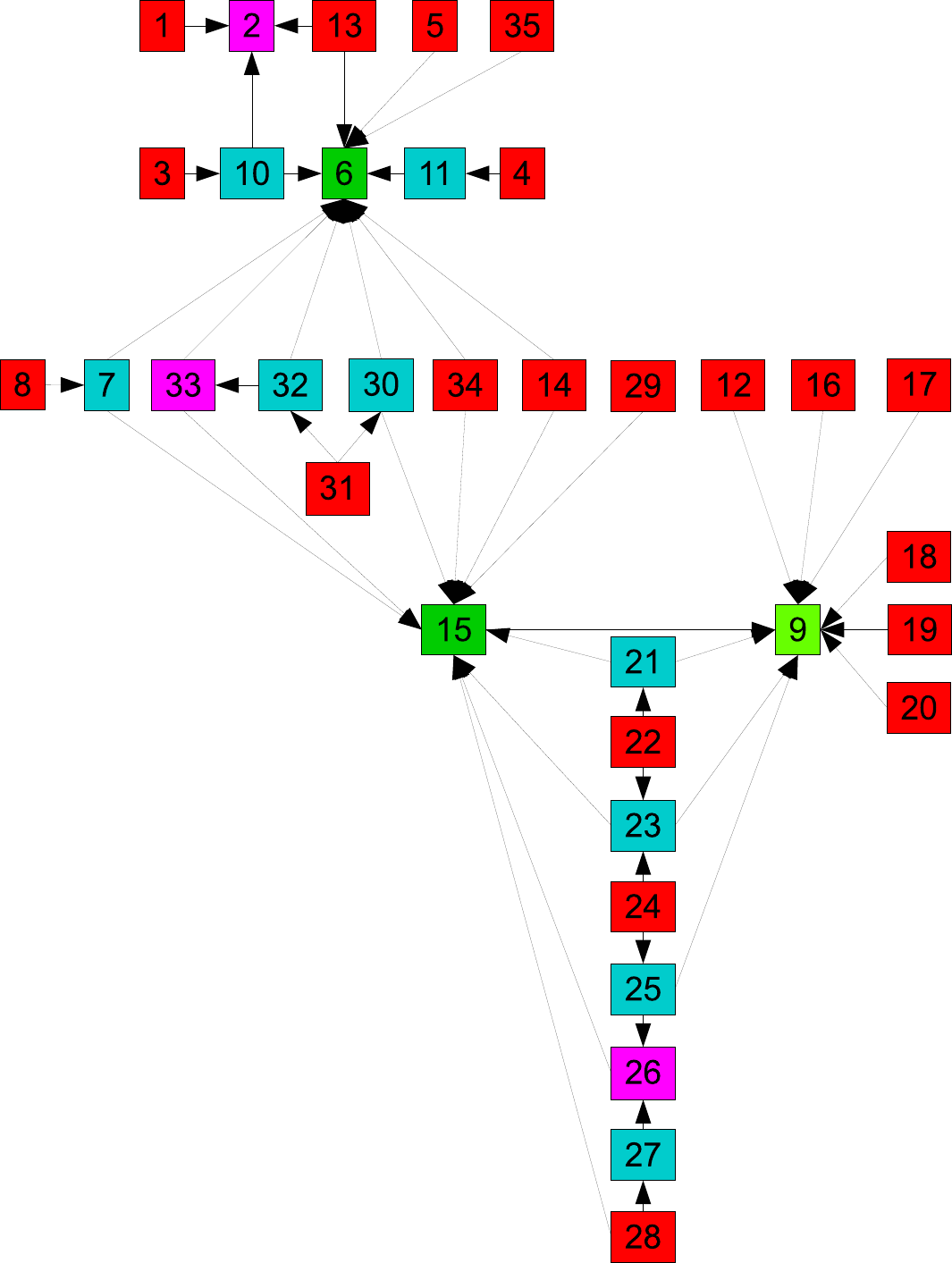}\\
 \caption{Graph corresponding to the hierarchical crack pattern presented in~\cite[Fig. 4a]{Yang2025a} along with the corresponding DAG, where leaves are shown in red, their parents in cyan, grandparents in magenta, great-grandparents in dark green, and the root in light green (cf. Fig.~\ref{fig:DAGmerged})}\label{fig:graph}
\end{figure*}

Figure~\ref{fig:graph} demonstrates a sample of the graph corresponding to an HCP along with the corresponding DAG. The HCP corresponds to that presented in~\cite[Fig. 4a]{Yang2025a}. Different shades in DAG correspond to different generations of cracks. Note that the planarity of the DAG here is incidental; DAGs are generally non-planar.

\begin{figure*}
 \centering
 \includegraphics[width=\textwidth]{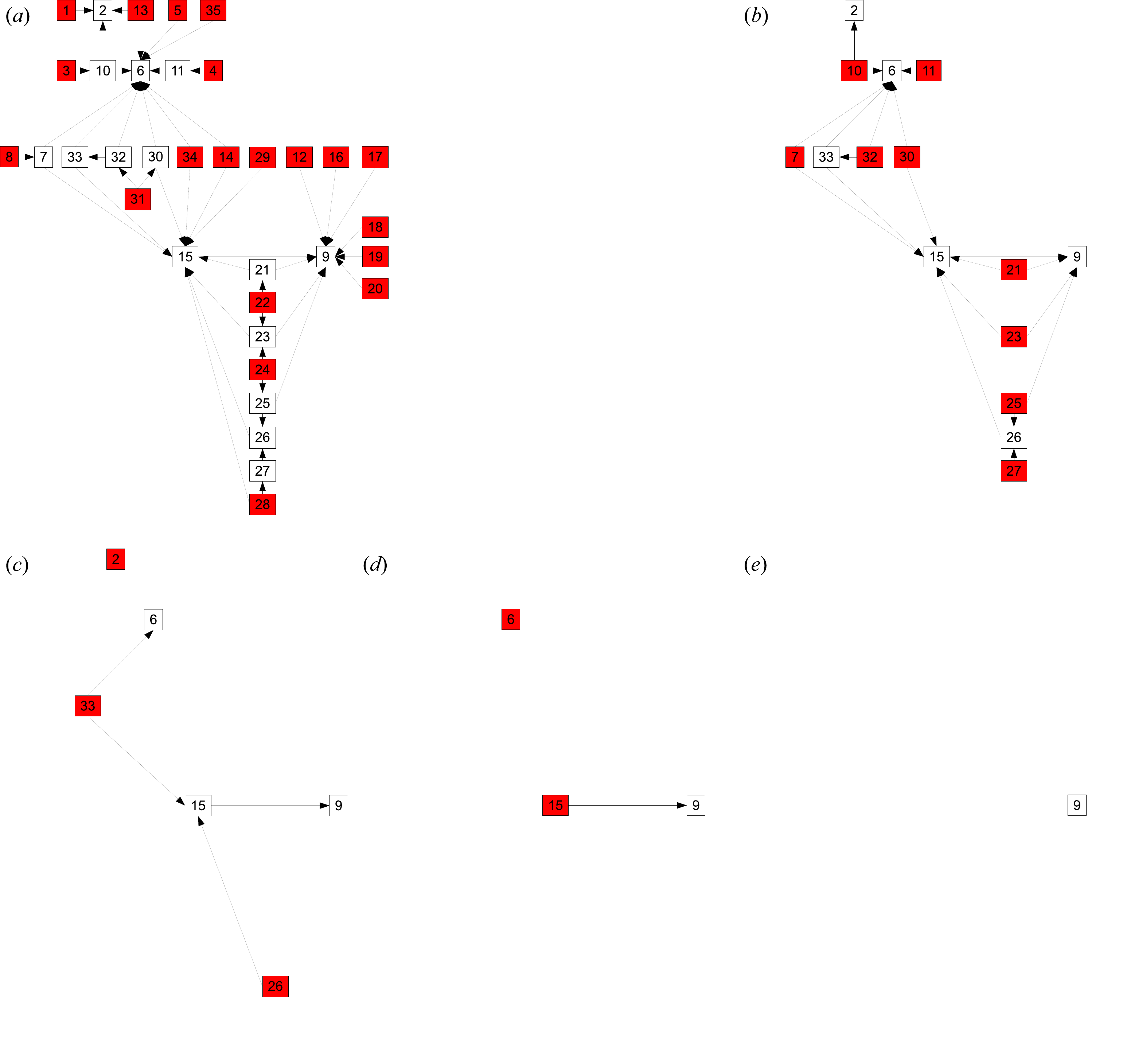}\\
 \caption{Stages of DAG processing. Highlighted vertices indicate cracks identified as leaves at each step.}\label{fig:DAGmerged}
\end{figure*}

Figure~\ref{fig:DAGmerged} presents stages of DAG processing.

\begin{figure*}
 \centering
 \includegraphics[width=\textwidth]{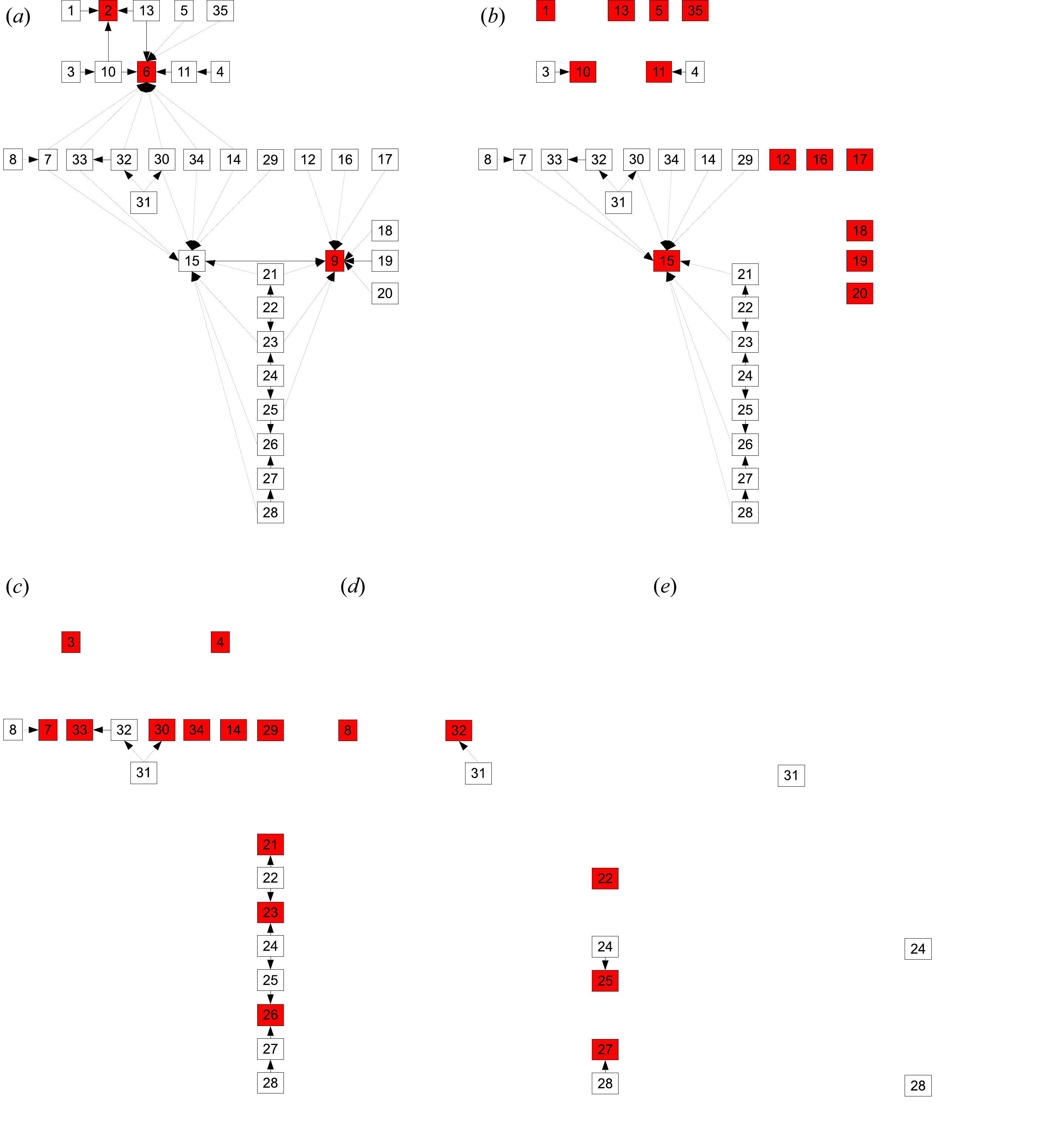}\\
 \caption{Stages of DAG processing in algorithm~\cite{Bohn2005}.}\label{fig:DAGBohnmerged}
\end{figure*}

For comparison, Fig.~\ref{fig:DAGBohnmerged} presents stages of DAG processing in algorithm~\cite{Bohn2005}.

To test the stability of the classification, two graphs were constructed that differed minimally from the original one. We used a color palette such that the leaves were colored the same in all cases. This coloring method is convenient because, generally speaking, the number of crack generations present in the image may change when the image boundary shifts. For comparison, it is convenient that the leaves retain their color even when the number of generations represented changes.

The classification exhibits high stability, even when the number of represented crack generations changes.
\begin{figure*}
 \centering
 \includegraphics[width=0.33\textwidth]{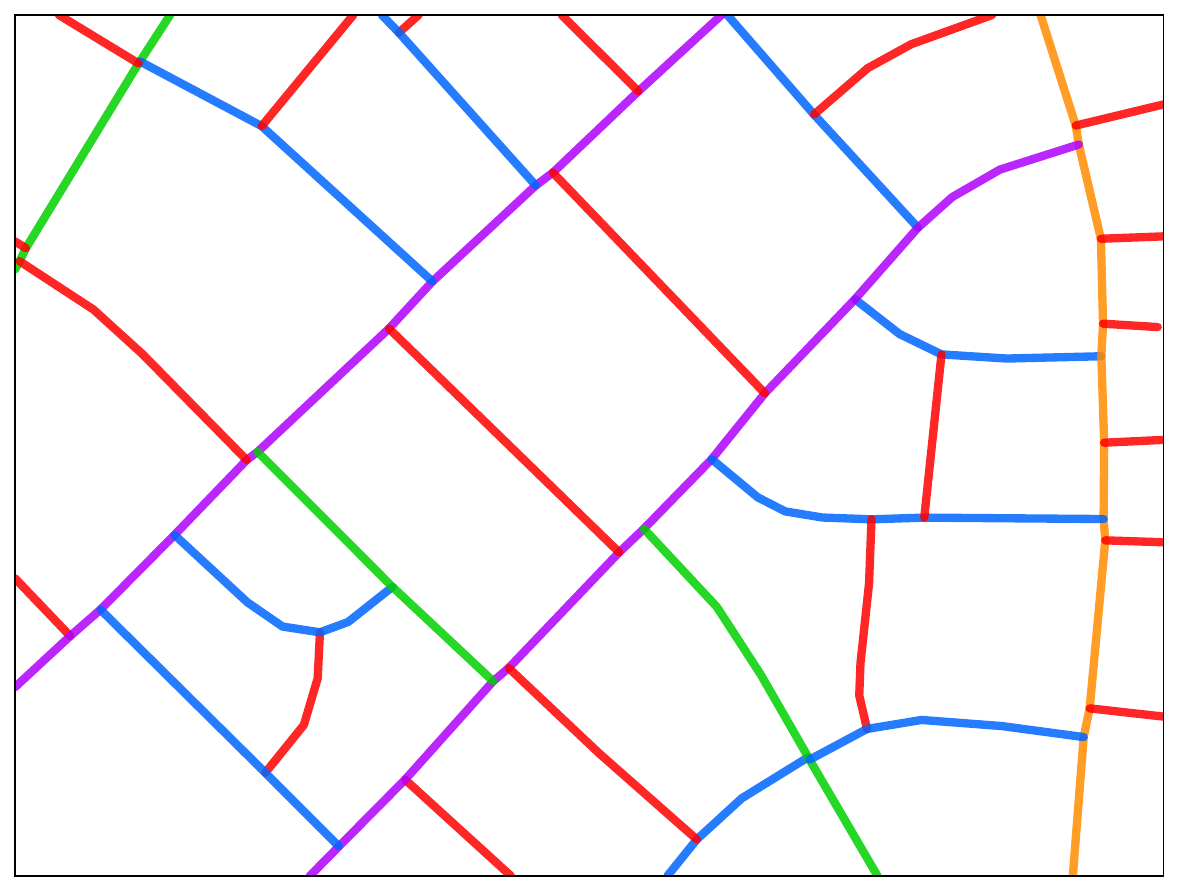}\hfill\includegraphics[width=0.33\textwidth]{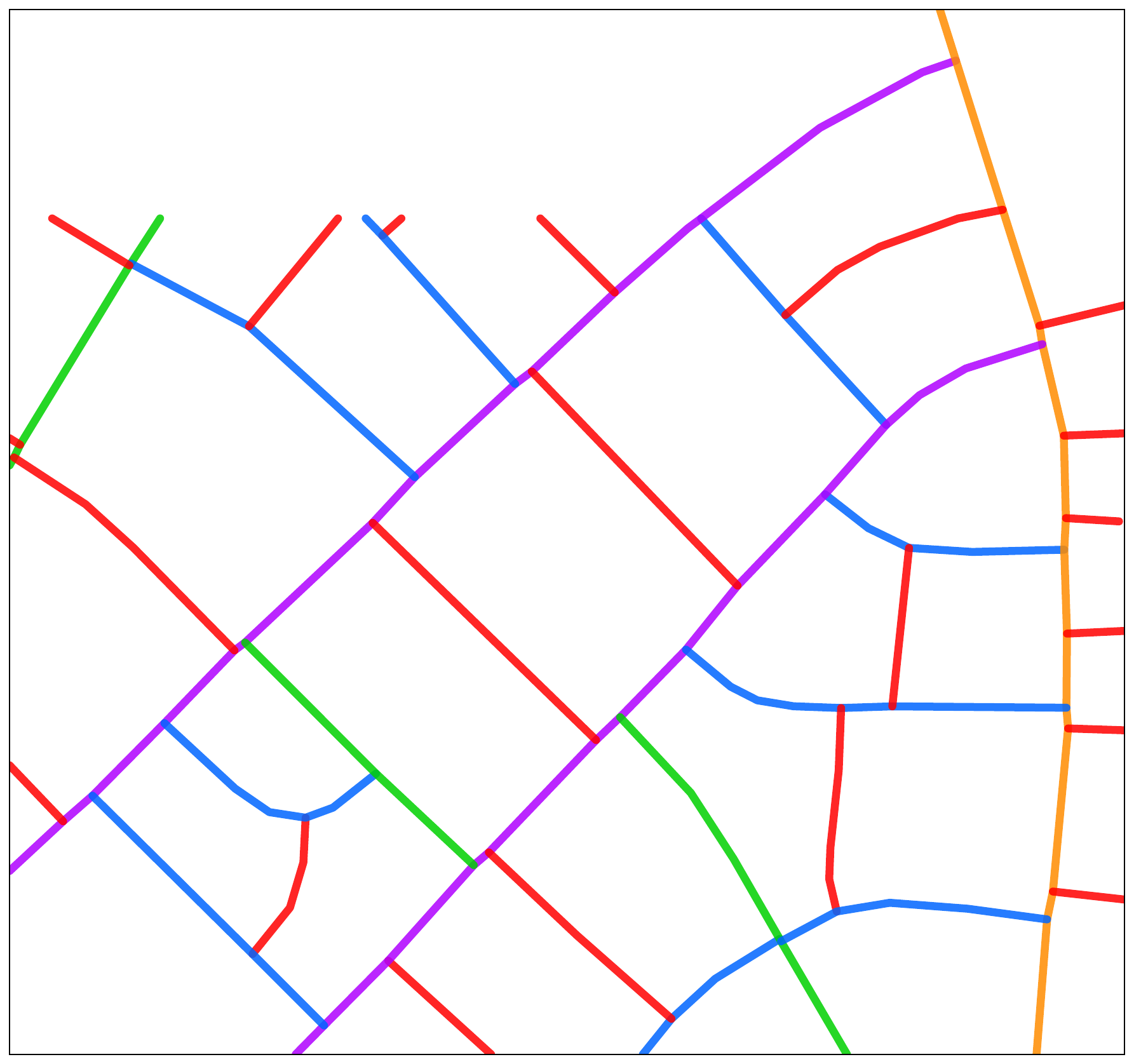}\hfill\includegraphics[width=0.33\textwidth]{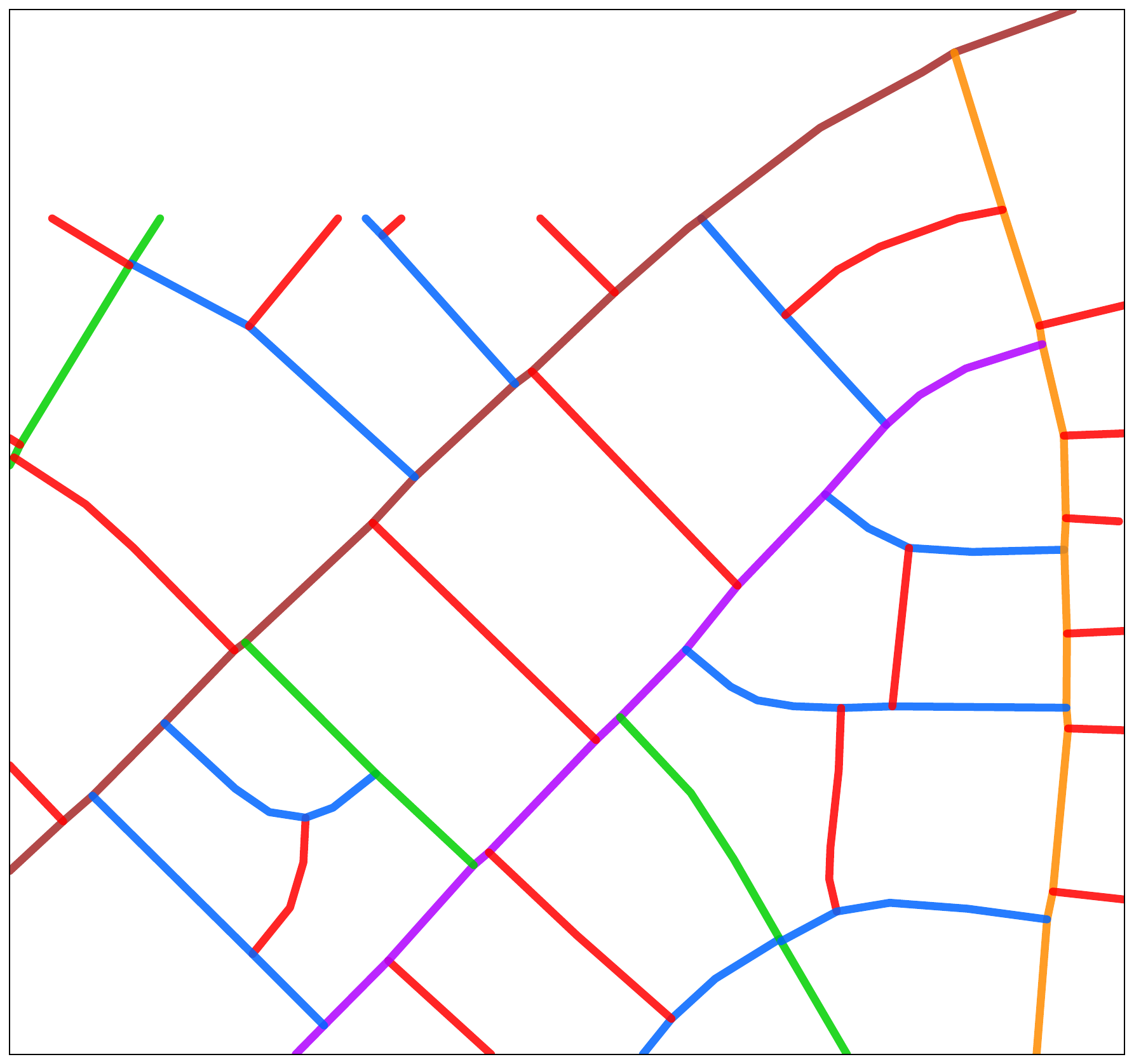}\\
 \caption{An example of the change in crack generations determined using DAG for two possible situations when shifting the image boundary. The different shades correspond to distances measured in generations from the leaves.}\label{fig:comparison}
\end{figure*}

\begin{figure*}
 \centering
\includegraphics[height=0.18\textheight]{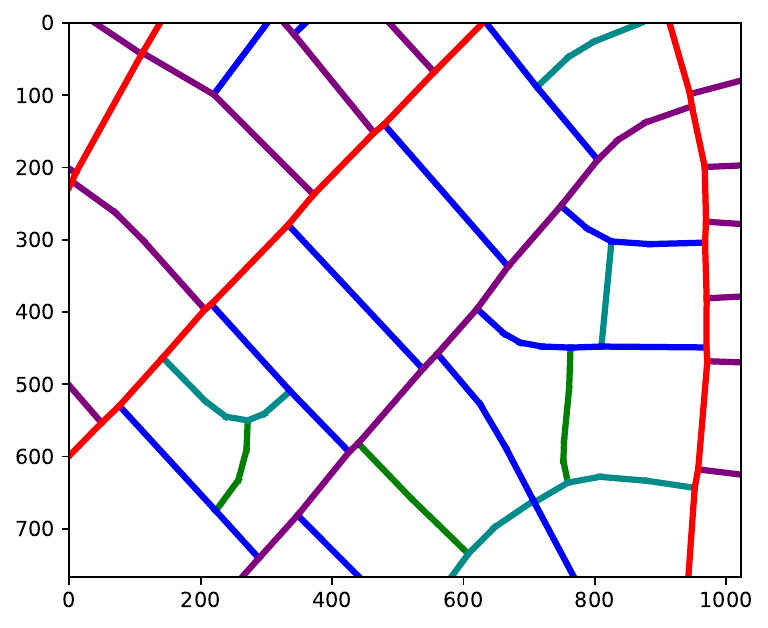}\includegraphics[height=0.21\textheight]{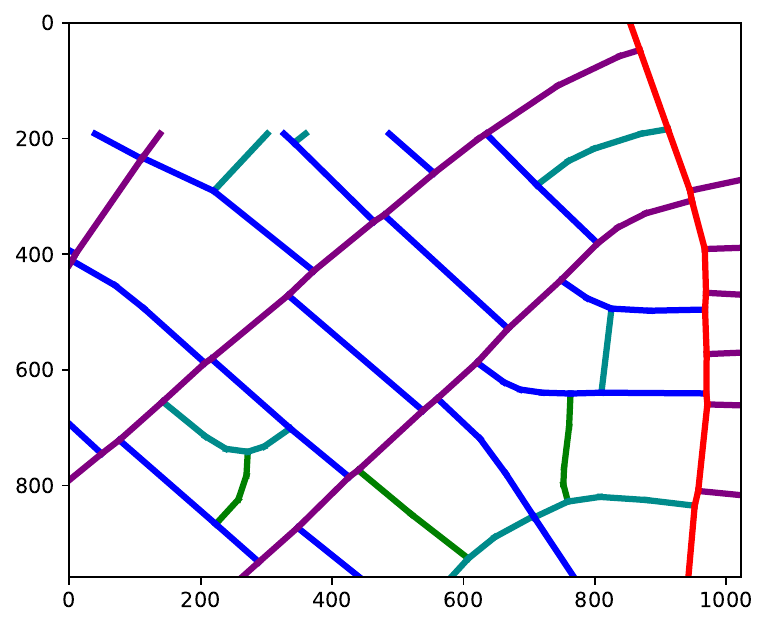}\includegraphics[height=0.21\textheight]{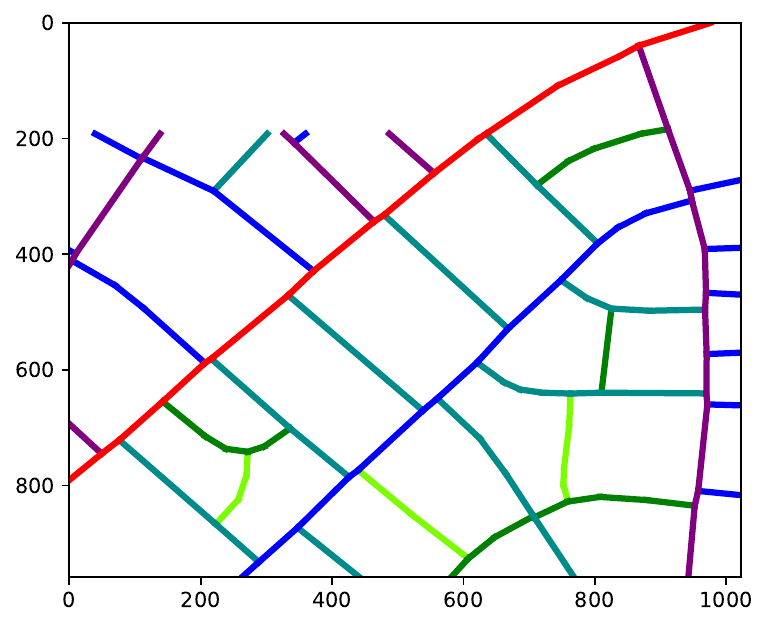}\\
 \caption{An example of the change in crack generations determined using algorithm~\cite{Bohn2005} for two possible situations when shifting the image boundary. Different shades correspond to different generations.}\label{fig:Bohn}
\end{figure*}

For comparison, Fig.~\ref{fig:Bohn} shows the results of the algorithm~\cite{Bohn2005} for the same situations. It is clear that the crack generation assignments change radically even with minor changes to the initial system.

To establish correspondences between cracks in the original image and those in its fragments, we construct a spatial index that maps each point's coordinates to the label of the crack passing through that point. A crack in a fragment is assigned the label of the original crack if it either completely coincides with that original crack or forms a part of it within the same coordinate system. Based on this matching, we estimate the stability of crack generation as the proportion of cracks in the fragment whose labels coincide with the labels in the original image.

\begin{figure}[!htbp]
  \centering
  \includegraphics[width=\columnwidth]{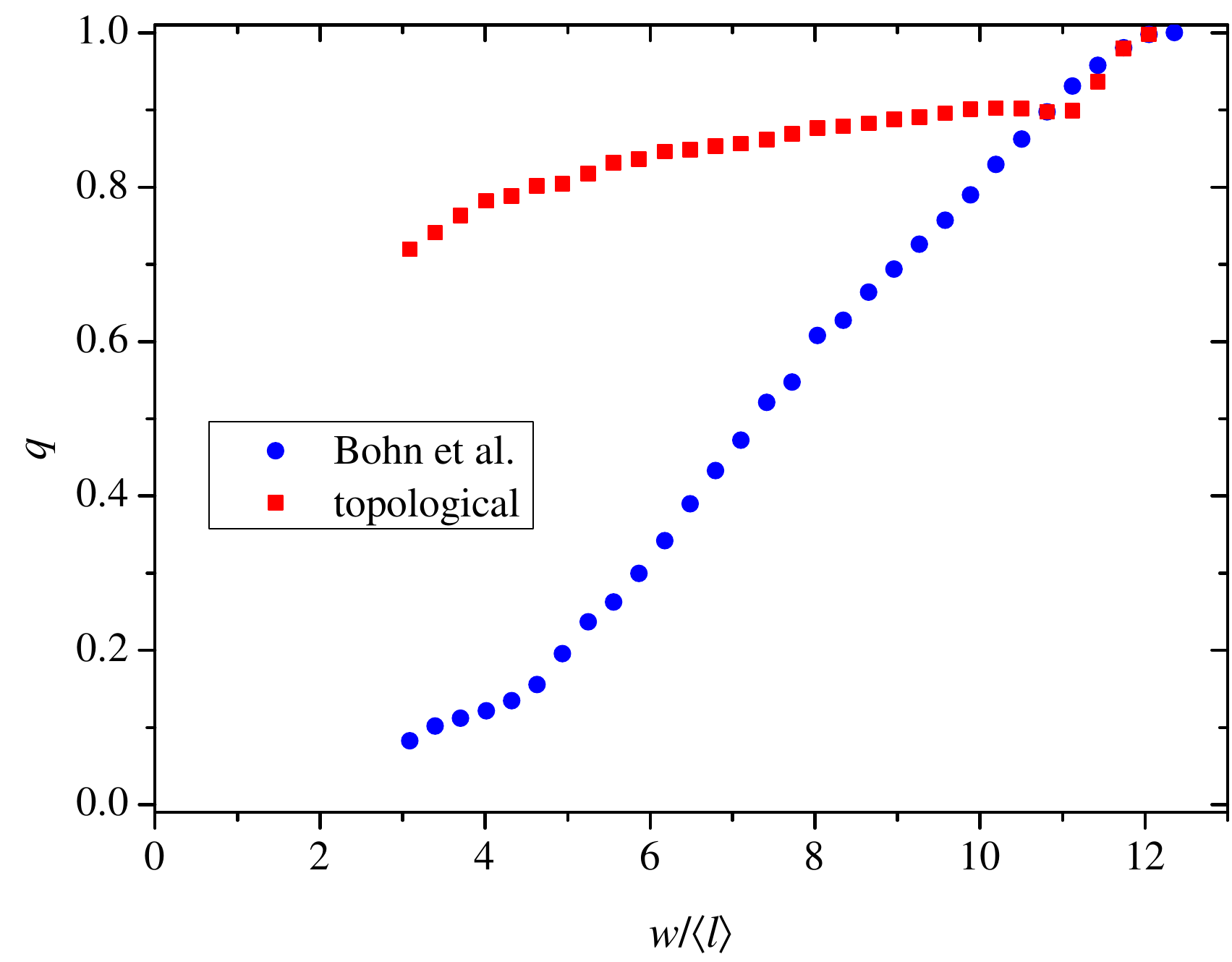}
  \caption{Plot of the fraction of correctly classified cracks, $q$, against the reciprocal window size (for two classification methods). The size of the markers is of the same order of magnitude as the standard error of the mean.}\label{fig:comparison1}
\end{figure}

Figure~\ref{fig:comparison1} compares stability of crack classifications with respect to image size. Results are averaged over 100 computer-generated samples obtained using the simulation algorithm described in Ref.~\cite{Tarasevich2026}. Squares marked as `Bohn et al.' correspond to the algorithm~\cite{Bohn2005}. Circles marked as `topological' correspond to the method described in this work.
As long as the window size, $w$, is greater than $\approx 12.5$ average edge length, $\langle l \rangle$, both classifications are accurate. As the relative window size decreases, the quality of classification~\cite{Bohn2005} decreases approximately linearly to zero. The quality of our classification also decreases, but significantly more slowly: even when the window size is only 3 times the average edge length, 70\% of cracks are correctly classified, while classification~\cite{Bohn2005} is correct for only 10\% of cracks.

\section{Conclusion\label{sec:concl}}

This study proposes a method for classifying crack generations in an image of a fragment of a hierarchical crack pattern. The problem is reduced to topological sorting of a directed acyclic graph whose vertices correspond to cracks, and whose arcs are directed from descendant to parent.

Graph theory was employed to extract edge chains and analyze the network topology; crack classification by generation was performed according to a method based on the hierarchy of crack junctions. Existing classification approaches were shown to have limitations related to the use of discrete time scale and images of only small fragments of the real-world crack patterns, which hinders the unambiguous determination of crack generations from final images.

Graph construction is based on the analysis of T-junctions and chains of edges that are close to a straight line, provided that the degree of vertices does not exceed three and there are no loops. An algorithm for sequentially removing leaves (vertices with zero in-degree) allows assigning numbers to generations, starting with the youngest.

Numerical experiments show that the proposed classification is highly stable to reasonable shifts in image boundaries: even with changes in the number of visible generations, the overall structure of the generations remains stable. Unlike an algorithm that begins classification with root cracks~\cite{Bohn2005}, our approach does not lead to radical restructuring with small shifts in the boundary of the accessible image of a network fragment.

Our study does not aim to address the influence of hierarchical structure on crack network properties or their applications. Rather, we focus on a specific methodological aspect: solving the generation-classification problem within such networks using the directed acyclic graph (DAG), a structure fundamental to mathematics and computer science. This approach serves as a bridge between the two disciplines.

The limitations of the method are related to the assumption of a strictly hierarchical network structure. Further research could be aimed at generalizing the algorithm to networks with more complex topologies and at incorporating additional information (such as crack widths) when available. However, a unified approach to handling the situations depicted in Figure 1 can be outlined. During preprocessing, loops and stars are readily identifiable. The cracks forming these structures should be marked, and their articulation points severed. Breaking these points is expected to transform the network into a strictly hierarchical one. Once crack generations are identified, it is necessary to verify that all cracks within a given loop or star belong to the same generation. Should this not be the case, a cyclical adjustment procedure is required.

Hierarchical networks are ubiquitous in the real world. Beyond crack patterns, prominent examples include leaf venation and urban transportation grids~\cite{Bohn2005a}. Yet the morphology of such networks varies widely: for instance, the venation of leaves exhibits a `hierarchically nested architecture'~\cite{Katifori2012}. In such networks, the `descendant--ancestor' relationship is not necessarily manifested as a T-junction, and the network hierarchy may be expressed, for example, not through a hierarchy of crack lengths and widths but through the nesting of loops. Thus, it is only natural that different algorithms should be applied to the analysis of hierarchical networks of different natures and, accordingly, different morphologies. For example, the 'hierarchical loop decomposition' algorithm was proposed for the analysis of hierarchical venation networks of plant leaves~\cite{Katifori2012}.

\begin{acknowledgments}
A.V.E. and A.S.B. acknowledge funding from the Russian Science Foundation, grant No.~25-21-00460.
\end{acknowledgments}

\clearpage
%
%

\bibliography{long,cracks}

\begin{thebibliography}{24}%
\makeatletter
\providecommand \@ifxundefined [1]{%
 \@ifx{#1\undefined}
}%
\providecommand \@ifnum [1]{%
 \ifnum #1\expandafter \@firstoftwo
 \else \expandafter \@secondoftwo
 \fi
}%
\providecommand \@ifx [1]{%
 \ifx #1\expandafter \@firstoftwo
 \else \expandafter \@secondoftwo
 \fi
}%
\providecommand \natexlab [1]{#1}%
\providecommand \enquote  [1]{``#1''}%
\providecommand \bibnamefont  [1]{#1}%
\providecommand \bibfnamefont [1]{#1}%
\providecommand \citenamefont [1]{#1}%
\providecommand \href@noop [0]{\@secondoftwo}%
\providecommand \href [0]{\begingroup \@sanitize@url \@href}%
\providecommand \@href[1]{\@@startlink{#1}\@@href}%
\providecommand \@@href[1]{\endgroup#1\@@endlink}%
\providecommand \@sanitize@url [0]{\catcode `\\12\catcode `\$12\catcode
  `\&12\catcode `\#12\catcode `\^12\catcode `\_12\catcode `\%12\relax}%
\providecommand \@@startlink[1]{}%
\providecommand \@@endlink[0]{}%
\providecommand \url  [0]{\begingroup\@sanitize@url \@url }%
\providecommand \@url [1]{\endgroup\@href {#1}{\urlprefix }}%
\providecommand \urlprefix  [0]{URL }%
\providecommand \Eprint [0]{\href }%
\providecommand \doibase [0]{https://doi.org/}%
\providecommand \selectlanguage [0]{\@gobble}%
\providecommand \bibinfo  [0]{\@secondoftwo}%
\providecommand \bibfield  [0]{\@secondoftwo}%
\providecommand \translation [1]{[#1]}%
\providecommand \BibitemOpen [0]{}%
\providecommand \bibitemStop [0]{}%
\providecommand \bibitemNoStop [0]{.\EOS\space}%
\providecommand \EOS [0]{\spacefactor3000\relax}%
\providecommand \BibitemShut  [1]{\csname bibitem#1\endcsname}%
\let\auto@bib@innerbib\@empty
\bibitem [{\citenamefont {Goehring}\ \emph {et~al.}(2015)\citenamefont
  {Goehring}, \citenamefont {Nakahara}, \citenamefont {Dutta}, \citenamefont
  {Kitsunezaki},\ and\ \citenamefont {Tarafdar}}]{Goehring2015}%
  \BibitemOpen
  \bibfield  {author} {\bibinfo {author} {\bibfnamefont {L.}~\bibnamefont
  {Goehring}}, \bibinfo {author} {\bibfnamefont {A.}~\bibnamefont {Nakahara}},
  \bibinfo {author} {\bibfnamefont {T.}~\bibnamefont {Dutta}}, \bibinfo
  {author} {\bibfnamefont {S.}~\bibnamefont {Kitsunezaki}},\ and\ \bibinfo
  {author} {\bibfnamefont {S.}~\bibnamefont {Tarafdar}},\ }\href
  {https://doi.org/10.1002/9783527671922} {\emph {\bibinfo {title} {Desiccation
  Cracks and their Patterns: Formation and Modelling in Science and Nature}}}\
  (\bibinfo  {publisher} {Wiley},\ \bibinfo {year} {2015})\BibitemShut
  {NoStop}%
\bibitem [{\citenamefont {Gupta}\ \emph {et~al.}(2025)\citenamefont {Gupta},
  \citenamefont {Sundaram},\ and\ \citenamefont {Kulkarni}}]{Gupta2025}%
  \BibitemOpen
  \bibfield  {author} {\bibinfo {author} {\bibfnamefont {R.}~\bibnamefont
  {Gupta}}, \bibinfo {author} {\bibfnamefont {K.~S.}\ \bibnamefont
  {Sundaram}},\ and\ \bibinfo {author} {\bibfnamefont {G.~U.}\ \bibnamefont
  {Kulkarni}},\ }\bibinfo {title} {Naturally ordered cracks},\ in\ \href
  {https://doi.org/10.1016/b978-0-443-15684-7.00005-1} {\emph {\bibinfo
  {booktitle} {Nature-Inspired Sensors}}}\ (\bibinfo  {publisher} {Elsevier},\
  \bibinfo {year} {2025})\ pp.\ \bibinfo {pages} {43--57}\BibitemShut {NoStop}%
\bibitem [{\citenamefont {Bohn}\ \emph
  {et~al.}(2005{\natexlab{a}})\citenamefont {Bohn}, \citenamefont {Pauchard},\
  and\ \citenamefont {Couder}}]{Bohn2005}%
  \BibitemOpen
  \bibfield  {author} {\bibinfo {author} {\bibfnamefont {S.}~\bibnamefont
  {Bohn}}, \bibinfo {author} {\bibfnamefont {L.}~\bibnamefont {Pauchard}},\
  and\ \bibinfo {author} {\bibfnamefont {Y.}~\bibnamefont {Couder}},\ }\href
  {https://doi.org/10.1103/PhysRevE.71.046214} {\bibfield  {journal} {\bibinfo
  {journal} {Phys. Rev. E}\ }\textbf {\bibinfo {volume} {71}},\ \bibinfo
  {pages} {046214} (\bibinfo {year} {2005}{\natexlab{a}})}\BibitemShut
  {NoStop}%
\bibitem [{\citenamefont {Schweiss}\ and\ \citenamefont
  {Pauchard}(2026)}]{Schweiss2026}%
  \BibitemOpen
  \bibfield  {author} {\bibinfo {author} {\bibfnamefont {T.}~\bibnamefont
  {Schweiss}}\ and\ \bibinfo {author} {\bibfnamefont {L.}~\bibnamefont
  {Pauchard}},\ }\href {https://doi.org/10.1103/93w4-6xc6} {\bibfield
  {journal} {\bibinfo  {journal} {Phys. Rev. E}\ }\textbf {\bibinfo {volume}
  {113}},\ \bibinfo {pages} {045503} (\bibinfo {year} {2026})}\BibitemShut
  {NoStop}%
\bibitem [{\citenamefont {Cama}\ \emph
  {et~al.}(2025{\natexlab{a}})\citenamefont {Cama}, \citenamefont {Pasini},
  \citenamefont {Giovanella},\ and\ \citenamefont {Galeotti}}]{Cama2025a}%
  \BibitemOpen
  \bibfield  {author} {\bibinfo {author} {\bibfnamefont {E.~S.}\ \bibnamefont
  {Cama}}, \bibinfo {author} {\bibfnamefont {M.}~\bibnamefont {Pasini}},
  \bibinfo {author} {\bibfnamefont {U.}~\bibnamefont {Giovanella}},\ and\
  \bibinfo {author} {\bibfnamefont {F.}~\bibnamefont {Galeotti}},\ }\href
  {https://doi.org/10.3390/coatings15020189} {\bibfield  {journal} {\bibinfo
  {journal} {Coatings}\ }\textbf {\bibinfo {volume} {15}},\ \bibinfo {pages}
  {189} (\bibinfo {year} {2025}{\natexlab{a}})}\BibitemShut {NoStop}%
\bibitem [{\citenamefont {Cama}\ \emph
  {et~al.}(2025{\natexlab{b}})\citenamefont {Cama}, \citenamefont {Pasini},
  \citenamefont {Galeotti},\ and\ \citenamefont {Giovanella}}]{Cama2025}%
  \BibitemOpen
  \bibfield  {author} {\bibinfo {author} {\bibfnamefont {E.~S.}\ \bibnamefont
  {Cama}}, \bibinfo {author} {\bibfnamefont {M.}~\bibnamefont {Pasini}},
  \bibinfo {author} {\bibfnamefont {F.}~\bibnamefont {Galeotti}},\ and\
  \bibinfo {author} {\bibfnamefont {U.}~\bibnamefont {Giovanella}},\ }\href
  {https://doi.org/10.3390/ma18133091} {\bibfield  {journal} {\bibinfo
  {journal} {Materials}\ }\textbf {\bibinfo {volume} {18}},\ \bibinfo {pages}
  {3091} (\bibinfo {year} {2025}{\natexlab{b}})}\BibitemShut {NoStop}%
\bibitem [{\citenamefont {Lazarus}\ and\ \citenamefont
  {Pauchard}(2011)}]{Lazarus2011}%
  \BibitemOpen
  \bibfield  {author} {\bibinfo {author} {\bibfnamefont {V.}~\bibnamefont
  {Lazarus}}\ and\ \bibinfo {author} {\bibfnamefont {L.}~\bibnamefont
  {Pauchard}},\ }\href {https://doi.org/10.1039/c0sm00900h} {\bibfield
  {journal} {\bibinfo  {journal} {Soft Matter}\ }\textbf {\bibinfo {volume}
  {7}},\ \bibinfo {pages} {2552} (\bibinfo {year} {2011})}\BibitemShut
  {NoStop}%
\bibitem [{\citenamefont {Goehring}(2013)}]{Goehring2013}%
  \BibitemOpen
  \bibfield  {author} {\bibinfo {author} {\bibfnamefont {L.}~\bibnamefont
  {Goehring}},\ }\href {https://doi.org/10.1098/rsta.2012.0353} {\bibfield
  {journal} {\bibinfo  {journal} {Philos. Trans. Royal Soc. A}\ }\textbf
  {\bibinfo {volume} {371}},\ \bibinfo {pages} {20120353} (\bibinfo {year}
  {2013})}\BibitemShut {NoStop}%
\bibitem [{\citenamefont {Pauchard}(2020)}]{Pauchard2020}%
  \BibitemOpen
  \bibfield  {author} {\bibinfo {author} {\bibfnamefont {L.}~\bibnamefont
  {Pauchard}},\ }\href {https://doi.org/10.5802/crmeca.33} {\bibfield
  {journal} {\bibinfo  {journal} {Comptes Rendus. M\'{e}canique}\ }\textbf
  {\bibinfo {volume} {348}},\ \bibinfo {pages} {637} (\bibinfo {year}
  {2020})}\BibitemShut {NoStop}%
\bibitem [{\citenamefont {Guan}\ \emph {et~al.}(2025)\citenamefont {Guan},
  \citenamefont {Yang}, \citenamefont {Chen}, \citenamefont {Wan},
  \citenamefont {Guo},\ and\ \citenamefont {Wang}}]{Guan2025}%
  \BibitemOpen
  \bibfield  {author} {\bibinfo {author} {\bibfnamefont {Y.}~\bibnamefont
  {Guan}}, \bibinfo {author} {\bibfnamefont {L.}~\bibnamefont {Yang}}, \bibinfo
  {author} {\bibfnamefont {C.}~\bibnamefont {Chen}}, \bibinfo {author}
  {\bibfnamefont {R.}~\bibnamefont {Wan}}, \bibinfo {author} {\bibfnamefont
  {C.}~\bibnamefont {Guo}},\ and\ \bibinfo {author} {\bibfnamefont
  {P.}~\bibnamefont {Wang}},\ }\href
  {https://doi.org/10.1016/j.isci.2024.111543} {\bibfield  {journal} {\bibinfo
  {journal} {iScience}\ }\textbf {\bibinfo {volume} {28}},\ \bibinfo {pages}
  {111543} (\bibinfo {year} {2025})}\BibitemShut {NoStop}%
\bibitem [{\citenamefont {Kranz}(1979)}]{Kranz1979}%
  \BibitemOpen
  \bibfield  {author} {\bibinfo {author} {\bibfnamefont {R.~L.}\ \bibnamefont
  {Kranz}},\ }\href {https://doi.org/10.1016/0148-9062(79)90773-3} {\bibfield
  {journal} {\bibinfo  {journal} {Int. J. Rock Mech. Min. Sci. \& Geomech.
  Abstr.}\ }\textbf {\bibinfo {volume} {16}},\ \bibinfo {pages} {37} (\bibinfo
  {year} {1979})}\BibitemShut {NoStop}%
\bibitem [{\citenamefont {Perna}\ \emph {et~al.}(2011)\citenamefont {Perna},
  \citenamefont {Kuntz},\ and\ \citenamefont {Douady}}]{Perna2011}%
  \BibitemOpen
  \bibfield  {author} {\bibinfo {author} {\bibfnamefont {A.}~\bibnamefont
  {Perna}}, \bibinfo {author} {\bibfnamefont {P.}~\bibnamefont {Kuntz}},\ and\
  \bibinfo {author} {\bibfnamefont {S.}~\bibnamefont {Douady}},\ }\href
  {https://doi.org/10.1103/physreve.83.066106} {\bibfield  {journal} {\bibinfo
  {journal} {Phys. Rev. E}\ }\textbf {\bibinfo {volume} {83}},\ \bibinfo
  {pages} {066106} (\bibinfo {year} {2011})}\BibitemShut {NoStop}%
\bibitem [{\citenamefont {Kumar}\ and\ \citenamefont
  {Kulkarni}(2021)}]{Kumar2021}%
  \BibitemOpen
  \bibfield  {author} {\bibinfo {author} {\bibfnamefont {A.}~\bibnamefont
  {Kumar}}\ and\ \bibinfo {author} {\bibfnamefont {G.~U.}\ \bibnamefont
  {Kulkarni}},\ }\href {https://doi.org/10.1021/acs.langmuir.1c02363}
  {\bibfield  {journal} {\bibinfo  {journal} {Langmuir}\ }\textbf {\bibinfo
  {volume} {37}},\ \bibinfo {pages} {13141} (\bibinfo {year}
  {2021})}\BibitemShut {NoStop}%
\bibitem [{\citenamefont {Voronin}\ \emph {et~al.}(2022)\citenamefont
  {Voronin}, \citenamefont {Fadeev}, \citenamefont {Makeev}, \citenamefont
  {Mikhalev}, \citenamefont {Osipkov}, \citenamefont {Provatorov},
  \citenamefont {Ryzhenko}, \citenamefont {Yurkov}, \citenamefont {Simunin},
  \citenamefont {Karpova}, \citenamefont {Lukyanenko}, \citenamefont {Kokh},
  \citenamefont {Bainov}, \citenamefont {Tambasov}, \citenamefont {Nedelin},
  \citenamefont {Zolotovsky},\ and\ \citenamefont {Khartov}}]{Voronin2022}%
  \BibitemOpen
  \bibfield  {author} {\bibinfo {author} {\bibfnamefont {A.~S.}\ \bibnamefont
  {Voronin}}, \bibinfo {author} {\bibfnamefont {Y.~V.}\ \bibnamefont {Fadeev}},
  \bibinfo {author} {\bibfnamefont {M.~O.}\ \bibnamefont {Makeev}}, \bibinfo
  {author} {\bibfnamefont {P.~A.}\ \bibnamefont {Mikhalev}}, \bibinfo {author}
  {\bibfnamefont {A.~S.}\ \bibnamefont {Osipkov}}, \bibinfo {author}
  {\bibfnamefont {A.~S.}\ \bibnamefont {Provatorov}}, \bibinfo {author}
  {\bibfnamefont {D.~S.}\ \bibnamefont {Ryzhenko}}, \bibinfo {author}
  {\bibfnamefont {G.~Y.}\ \bibnamefont {Yurkov}}, \bibinfo {author}
  {\bibfnamefont {M.~M.}\ \bibnamefont {Simunin}}, \bibinfo {author}
  {\bibfnamefont {D.~V.}\ \bibnamefont {Karpova}}, \bibinfo {author}
  {\bibfnamefont {A.~V.}\ \bibnamefont {Lukyanenko}}, \bibinfo {author}
  {\bibfnamefont {D.}~\bibnamefont {Kokh}}, \bibinfo {author} {\bibfnamefont
  {D.~D.}\ \bibnamefont {Bainov}}, \bibinfo {author} {\bibfnamefont {I.~A.}\
  \bibnamefont {Tambasov}}, \bibinfo {author} {\bibfnamefont {S.~V.}\
  \bibnamefont {Nedelin}}, \bibinfo {author} {\bibfnamefont {N.~A.}\
  \bibnamefont {Zolotovsky}},\ and\ \bibinfo {author} {\bibfnamefont {S.~V.}\
  \bibnamefont {Khartov}},\ }\href {https://doi.org/10.3390/ma15041449}
  {\bibfield  {journal} {\bibinfo  {journal} {Materials}\ }\textbf {\bibinfo
  {volume} {15}},\ \bibinfo {pages} {1449} (\bibinfo {year}
  {2022})}\BibitemShut {NoStop}%
\bibitem [{\citenamefont {Fanfoni}\ \emph {et~al.}(2025)\citenamefont
  {Fanfoni}, \citenamefont {Bonanni}, \citenamefont {Addessi}, \citenamefont
  {Martini}, \citenamefont {Goletti},\ and\ \citenamefont
  {Sgarlata}}]{Fanfoni2025}%
  \BibitemOpen
  \bibfield  {author} {\bibinfo {author} {\bibfnamefont {M.}~\bibnamefont
  {Fanfoni}}, \bibinfo {author} {\bibfnamefont {B.}~\bibnamefont {Bonanni}},
  \bibinfo {author} {\bibfnamefont {S.}~\bibnamefont {Addessi}}, \bibinfo
  {author} {\bibfnamefont {R.}~\bibnamefont {Martini}}, \bibinfo {author}
  {\bibfnamefont {C.}~\bibnamefont {Goletti}},\ and\ \bibinfo {author}
  {\bibfnamefont {A.}~\bibnamefont {Sgarlata}},\ }\href
  {https://doi.org/10.1103/physreve.111.044127} {\bibfield  {journal} {\bibinfo
   {journal} {Phys. Rev. E}\ }\textbf {\bibinfo {volume} {111}},\ \bibinfo
  {pages} {044127} (\bibinfo {year} {2025})}\BibitemShut {NoStop}%
\bibitem [{\citenamefont {Nahlawi}\ and\ \citenamefont
  {Kodikara}(2006)}]{Nahlawi2006}%
  \BibitemOpen
  \bibfield  {author} {\bibinfo {author} {\bibfnamefont {H.}~\bibnamefont
  {Nahlawi}}\ and\ \bibinfo {author} {\bibfnamefont {J.~K.}\ \bibnamefont
  {Kodikara}},\ }\href {https://doi.org/10.1007/s10706-005-4894-4} {\bibfield
  {journal} {\bibinfo  {journal} {Geotech. Geol. Eng.}\ }\textbf {\bibinfo
  {volume} {24}},\ \bibinfo {pages} {1641} (\bibinfo {year}
  {2006})}\BibitemShut {NoStop}%
\bibitem [{\citenamefont {Tang}\ \emph {et~al.}(2011)\citenamefont {Tang},
  \citenamefont {Shi}, \citenamefont {Liu}, \citenamefont {Suo},\ and\
  \citenamefont {Gao}}]{Tang2011}%
  \BibitemOpen
  \bibfield  {author} {\bibinfo {author} {\bibfnamefont {C.-S.}\ \bibnamefont
  {Tang}}, \bibinfo {author} {\bibfnamefont {B.}~\bibnamefont {Shi}}, \bibinfo
  {author} {\bibfnamefont {C.}~\bibnamefont {Liu}}, \bibinfo {author}
  {\bibfnamefont {W.-B.}\ \bibnamefont {Suo}},\ and\ \bibinfo {author}
  {\bibfnamefont {L.}~\bibnamefont {Gao}},\ }\href
  {https://doi.org/10.1016/j.clay.2011.01.032} {\bibfield  {journal} {\bibinfo
  {journal} {Appl. Clay Sci.}\ }\textbf {\bibinfo {volume} {52}},\ \bibinfo
  {pages} {69} (\bibinfo {year} {2011})}\BibitemShut {NoStop}%
\bibitem [{\citenamefont {Tian}\ \emph {et~al.}(2023)\citenamefont {Tian},
  \citenamefont {Cheng}, \citenamefont {Tang},\ and\ \citenamefont
  {Shi}}]{Tian2023}%
  \BibitemOpen
  \bibfield  {author} {\bibinfo {author} {\bibfnamefont {B.-G.}\ \bibnamefont
  {Tian}}, \bibinfo {author} {\bibfnamefont {Q.}~\bibnamefont {Cheng}},
  \bibinfo {author} {\bibfnamefont {C.-S.}\ \bibnamefont {Tang}},\ and\
  \bibinfo {author} {\bibfnamefont {B.}~\bibnamefont {Shi}},\ }\href
  {https://doi.org/10.1016/j.enggeo.2022.106973} {\bibfield  {journal}
  {\bibinfo  {journal} {Eng. Geol.}\ }\textbf {\bibinfo {volume} {313}},\
  \bibinfo {pages} {106973} (\bibinfo {year} {2023})}\BibitemShut {NoStop}%
\bibitem [{\citenamefont {Yang}\ \emph
  {et~al.}(2025{\natexlab{a}})\citenamefont {Yang}, \citenamefont {Tang},
  \citenamefont {Wang}, \citenamefont {Cheng}, \citenamefont {Liu},
  \citenamefont {Zeng},\ and\ \citenamefont {Shen}}]{Yang2025b}%
  \BibitemOpen
  \bibfield  {author} {\bibinfo {author} {\bibfnamefont {Z.-M.}\ \bibnamefont
  {Yang}}, \bibinfo {author} {\bibfnamefont {C.-S.}\ \bibnamefont {Tang}},
  \bibinfo {author} {\bibfnamefont {T.}~\bibnamefont {Wang}}, \bibinfo {author}
  {\bibfnamefont {Q.}~\bibnamefont {Cheng}}, \bibinfo {author} {\bibfnamefont
  {J.-D.}\ \bibnamefont {Liu}}, \bibinfo {author} {\bibfnamefont {Z.-X.}\
  \bibnamefont {Zeng}},\ and\ \bibinfo {author} {\bibfnamefont
  {Z.}~\bibnamefont {Shen}},\ }\href
  {https://doi.org/10.1016/j.enggeo.2025.108122} {\bibfield  {journal}
  {\bibinfo  {journal} {Eng. Geol.}\ }\textbf {\bibinfo {volume} {353}},\
  \bibinfo {pages} {108122} (\bibinfo {year} {2025}{\natexlab{a}})}\BibitemShut
  {NoStop}%
\bibitem [{\citenamefont {Kahn}(1962)}]{Kahn1962}%
  \BibitemOpen
  \bibfield  {author} {\bibinfo {author} {\bibfnamefont {A.~B.}\ \bibnamefont
  {Kahn}},\ }\href {https://doi.org/10.1145/368996.369025} {\bibfield
  {journal} {\bibinfo  {journal} {Commun. ACM}\ }\textbf {\bibinfo {volume}
  {5}},\ \bibinfo {pages} {558} (\bibinfo {year} {1962})}\BibitemShut {NoStop}%
\bibitem [{\citenamefont {Tarasevich}\ \emph {et~al.}(2026)\citenamefont
  {Tarasevich}, \citenamefont {Eserkepov},\ and\ \citenamefont
  {Burmistrov}}]{Tarasevich2026}%
  \BibitemOpen
  \bibfield  {author} {\bibinfo {author} {\bibfnamefont {Y.~Y.}\ \bibnamefont
  {Tarasevich}}, \bibinfo {author} {\bibfnamefont {A.~V.}\ \bibnamefont
  {Eserkepov}},\ and\ \bibinfo {author} {\bibfnamefont {A.~S.}\ \bibnamefont
  {Burmistrov}},\ }\href {https://doi.org/10.1103/jyvg-9yrl} {\bibfield
  {journal} {\bibinfo  {journal} {Phys. Rev. E}\ }\textbf {\bibinfo {volume}
  {114}},\ \bibinfo {pages} {014314} (\bibinfo {year} {2026})}\BibitemShut
  {NoStop}%
\bibitem [{\citenamefont {Yang}\ \emph
  {et~al.}(2025{\natexlab{b}})\citenamefont {Yang}, \citenamefont {Guo},
  \citenamefont {Gao}, \citenamefont {Guan}, \citenamefont {Zhang},\ and\
  \citenamefont {Wang}}]{Yang2025a}%
  \BibitemOpen
  \bibfield  {author} {\bibinfo {author} {\bibfnamefont {L.}~\bibnamefont
  {Yang}}, \bibinfo {author} {\bibfnamefont {R.}~\bibnamefont {Guo}}, \bibinfo
  {author} {\bibfnamefont {F.}~\bibnamefont {Gao}}, \bibinfo {author}
  {\bibfnamefont {Y.}~\bibnamefont {Guan}}, \bibinfo {author} {\bibfnamefont
  {M.}~\bibnamefont {Zhang}},\ and\ \bibinfo {author} {\bibfnamefont
  {P.}~\bibnamefont {Wang}},\ }\href {https://doi.org/10.3390/ma18051067}
  {\bibfield  {journal} {\bibinfo  {journal} {Materials}\ }\textbf {\bibinfo
  {volume} {18}},\ \bibinfo {pages} {1067} (\bibinfo {year}
  {2025}{\natexlab{b}})}\BibitemShut {NoStop}%
\bibitem [{\citenamefont {Bohn}\ \emph
  {et~al.}(2005{\natexlab{b}})\citenamefont {Bohn}, \citenamefont {Douady},\
  and\ \citenamefont {Couder}}]{Bohn2005a}%
  \BibitemOpen
  \bibfield  {author} {\bibinfo {author} {\bibfnamefont {S.}~\bibnamefont
  {Bohn}}, \bibinfo {author} {\bibfnamefont {S.}~\bibnamefont {Douady}},\ and\
  \bibinfo {author} {\bibfnamefont {Y.}~\bibnamefont {Couder}},\ }\href
  {https://doi.org/10.1103/physrevlett.94.054503} {\bibfield  {journal}
  {\bibinfo  {journal} {Phys. Rev. Lett.}\ }\textbf {\bibinfo {volume} {94}},\
  \bibinfo {pages} {054503} (\bibinfo {year} {2005}{\natexlab{b}})}\BibitemShut
  {NoStop}%
\bibitem [{\citenamefont {Katifori}\ and\ \citenamefont
  {Magnasco}(2012)}]{Katifori2012}%
  \BibitemOpen
  \bibfield  {author} {\bibinfo {author} {\bibfnamefont {E.}~\bibnamefont
  {Katifori}}\ and\ \bibinfo {author} {\bibfnamefont {M.~O.}\ \bibnamefont
  {Magnasco}},\ }\href {https://doi.org/10.1371/journal.pone.0037994}
  {\bibfield  {journal} {\bibinfo  {journal} {PLoS ONE}\ }\textbf {\bibinfo
  {volume} {7}},\ \bibinfo {pages} {e37994} (\bibinfo {year}
  {2012})}\BibitemShut {NoStop}%
\end{thebibliography}%

\end{document}